\documentclass[twocolumn,showpacs,preprintnumbers,amsmath,amssymb]{revtex4}
\usepackage{graphicx}
\usepackage{dcolumn}
\usepackage{bm}

\begin{document}

%\preprint{APS/123-QED}

\title{Singlet Superconductivity in Carbon Nanotubes}

\author{Kenji Kamide${}^1$}
\email{kamide@kh.phys.waseda.ac.jp}
\author{Takashi Kimura${}^{2,3}$}
\author{Munehiro Nishida${}^1$}
\author{Susumu Kurihara${}^1$}
\affiliation{${}^1$Department of Physics, Waseda University, \=Okubo, Shinjuku, Tokyo, 169-8555, Japan \\
${}^2$Advanced Research Institute for Science and Engineering, Waseda University, \=Okubo, Shinjuku, Tokyo, 169-8555, Japan \\
${}^3$Department of Physics, University of Tokyo, Hongo, Tokyo, 113-0033, Japan}
\date{\today}

\begin{abstract}
We discuss phase diagrams obtained by calculating temperature dependent correlation functions for superconductivity and density waves for a single $(5,0)$ carbon nanotube (CNs).
 We use one-loop renormalization group method within logarithmic accuracy.
In this system we must specify scattering channels in terms of momentum along the circumferential direction as well as the axis direction, since $(5,0)$ CN has two degenerate bands crossing the Fermi energy with circumferential momenta. 
We find singlet-superconducting phase with $5\hbar$ circumferential angular momentum.
 Critical exponents for each correlation functions are also calculated.
\end{abstract}

\maketitle

\section{\label{sec:level1}Introduction}
%\protect \lowercase{} \textbackslash\textbackslash}

%%%%%%%%%%%%%%%%%%%%%%%%%%%%%%%%%%%%%%%%%%%%%%%%%%%%%%%%%%%%%%%%%%%%%%%%%%%%%%%%%%%%%%%%%%%%%%%%% %%%%%%%%%%%%%%%%%%%%%%%%%%%%%%%%%%%%%%%%%%%%%%%%%%%%%%%%%%%%%%%%%%%%%%%%%%%%%%%%%%%%%%%%%%%%%%%%%%%%%%%%%%%%%%%%%%%%%%%%%%%%%%%%%%%%%%%%%%%%%%%%%%%%%

The discovery of carbon nanotubes (CN) in 1991~\cite{rf:1} has attracted much attention because of its potential for new physics as well as applications in electronic devices~\cite{rf:2}.
 A single-wall CN (SWCN) is made of a graphite layer rolled up into a cylinder with small diameter.
 SWCNs may be regarded as one-dimensional (1D) electron systems due to quantization of the circumferential momentum, as confirmed by experiments~\cite{rf:3}.

Recently, Tang $et$ $al.$~\cite{rf:4} reported evidence for superconductivity in a SWCN system below 15K. This system consists of $(5,0)$ CN with diameter 4\AA, separated from each other by zeolite walls.  
So this experiment implies that well-developed superconducting correlation (SCC) can exist in each individual nanotube.
The microscopic origin of the SCC in a SWCN is an interesting question to ask, since divergent 1D-SCC will eventually cause three-dimensional superconductivity with electron tunneling through zeolite walls. 

On the other hand, bosonization~\cite{rf:5,rf:6} and renormalization group~\cite{rf:7,rf:8} methods have been applied to the $(n,n)$CN to explain the superconductivity in the ropes of $(n,n)$CNs~\cite{rf:9}.
These works have shown that a singlet superconducting phase is more favored if short-range attractive interactions exist between electrons.
Moreover, some works ~\cite{rf:10,rf:11,rf:12} have suggested the presence of short-range attractive interactions in $(n,n)$CN.
However, results of bosonization and renormalization group~\cite{rf:5,rf:6,rf:7,rf:8} cannot be directly applied to $(5,0)$CN case, because the band structure of $(5,0)$CN is very different from that of $(n,n)$CN. 

Within the local-density approximation, (5,0) CN has three bands crossing the Fermi energy~\cite{rf:13}; one non-degenerate band and two two-fold-degenerate bands with opposite circumferential momenta, as shown in Fig.1.
 Such a band structure cannot be obtained within the tight-binding approximation because $\sigma$-$\pi$ hybridization effects change it significantly in a SWCN with diameter 4\AA~\cite{rf:14}. 
This band structure will provide opportunities to develop SCC with total circumferential momentum.
 We should also take into account the interaction between electrons with momentum transfer in the circumferential direction as well as in the axis direction. \par
In this paper, we perform one-loop RG calculation and derive the scaling equations for the three-band system, including the interactions with angular momentum transfer. 
We find phase diagrams in terms of several coupling constants by solving these equations numerically.
 Critical exponents which govern the temperature dependence of correlation functions near $T_{c}$ are also calculated.
%%%%%%%%%%%%%%%%%%%%%%%%%%%%%%%%%%%%%%%%%%%%%%%%%%%%%%%%%%%%%%%%%%%%%%%%%%%%%%%%%%%%%%%%%%%%%%  THREE BAND MODEL AND THREE CHANNELS   %%%%%%%%%%%%%%%%%%%%%%%%%%%%%%%%%%%%%%%%%%%%%%%%%%%%%%%%%%%%%%%%%%%%%%%%%%%%%%%%%%%%%%%%%%%%%%%%%%%%%%%%
\section{Three-band model and three groups}
We consider a system consisting of three linear bands near the Fermi level (Fig.1).
Band $0$ has no angular momentum $(\text{L}_{0}=0)$, band $1$ and $2$ have a finite angular momentum and are two-fold degenerate $(\text{L}_{1,2}=\pm\hbar n_{1,2})$.
In general two-particle correlation (TPC) are formed between branches with different sign in Fermi velocity in 1D electron systems.
 As we have three bands, six channels of TPC are produced.
We take forward and backward scatterings into account in this three-band system and perform g-ology~\cite{rf:15}, and determine the most divergent TPC among all possible ones by solving scaling equations.
%%%%%%%%%%%%%%%%%%%%%%%%%%%%%%%%%%%%%%%%%%%%%%%%%%%%%%%%%%%%%%%%%%%%%%%%%%%%%%%%%%%%%%%%%%%%%%%           FREE HAMILTONIAN           %%%%%%%%%%%%%%%%%%%%%%%%%%%%%%%%%%%%%%%%%%%%%%%%%%%%%%%%%%%%%%%%%%%%%%%%%%%%%%%%%%%%%%%%%%%%%%%%%%%%%%%%%%
We start from a Hamiltonian for non-interacting electrons:
\begin{eqnarray}
H_{0}=\sum_{\tau,\sigma,k}
{\tau v_{F,0} (k-\tau k_{F,0}) \ \psi^{\dagger}_{0,\tau,\sigma}(k) \psi_{0,\tau,\sigma}(k)} && \nonumber \\
-\sum_{
\begin{subarray}
x\gamma=1,2 \\
\tau,l,\sigma,k 
\end{subarray}
}
{\tau v_{F,\gamma} (k-\tau k_{F,\gamma}) \ \psi^{\dagger}_{\gamma,\tau,l,\sigma}(k) \psi_{\gamma,\tau,l,\sigma}(k)}, &&
\end{eqnarray}
\noindent
where $v_{F,\gamma}$ and $k_{F,\gamma}$ represent Fermi velocity and Fermi momentum for band $\gamma$.
 Spin, sign in Fermi momentum and in angular momentum $\text{L}_\gamma$ are denoted by $\sigma$, $\tau$ and $l=\pm$ respectively.
 We carry out a perturbation calculation with the interaction between electrons,
\begin{eqnarray}
H_{I}=\sum
g_{\gamma_1,\gamma_2,\gamma_3,\gamma_4}^{
\begin{subarray}
x\tau_1,\tau_2,\tau_3,\tau_4 \\
l_1,l_2,l_3,l_4
\end{subarray}
}
&&\psi^{\dagger}_{\gamma_1,\tau_1,l_1,\sigma}(k+q) 
\psi^{\dagger}_{\gamma_2,\tau_2,l_2,\sigma'}(k'-q) \nonumber \\
\times&&\psi_{\gamma_3,\tau_3,l_3,\sigma'}(k')
\psi_{\gamma_4,\tau_4,l_4,\sigma}(k).
\end{eqnarray}
Keeping the momentum transfer in mind both in the circumferential direction as well as in the axial direction, we abbreviate interaction constants as $g_i^j$
 where i and j represent the types of scattering processes in axial and circumferential direction respectively:
 $1=$backward, $2=$interbranch forward, $3=$Umklapp and $4=$intrabranch forward.
 We neglect the intra-branch forward scatterings $g_{4}^{j}$ because they are irrelevant within the one-loop RG framework.
 We also neglect Umklapp processes $g_{3}^{j}$ because (5,0) CN systems are far from half-filling for any band.
%%%%%%%%%%%%%%%%%%%%%%%%%%%%%%%%%%%%%%%%%%%%%%%%%%%%%%%%%%%%%%%%%%%%%%%%%%%%%%%%%%%%%%%%%%%%%%%    (5,0) CN Band Figure     %%%%%%%%%%%%%%%%%%%%%%%%%%%%%%%%%%%%%%%%%%%%%%%%%%%%%%%%%%%%%%%%%%%%%%%%%%%%%%%%%%%%%%%%%%%%%%%%%%%%%%%%%%%%%%%%%%%%%
\begin{figure}[b]
\includegraphics[height=3cm,width=5cm]{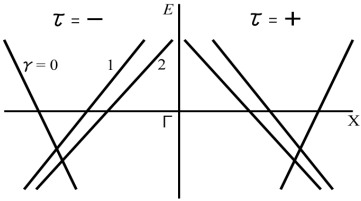}
\caption{\label{fig:epsart}
Band structure of (5,0) CN. Electrons are labeled by band $\gamma \in \{0,1,2\}$,
 Fermi point $\tau=\pm$, angular momentum $L_{\gamma}=\pm\hbar|n_\gamma|$ and spin $\sigma=\pm$.}
\end{figure}
 
 We divide six channels into three groups by considering the types of scattering processes.
 We denote a channel from band $\gamma$ and band $\gamma'$ by $(\gamma,\gamma')$.
%%%%%%%%%%%%%%%%%%%%%%%%%%%%%%%%%%%%%%%%%%%%%%%%%%%%%%%%%%%%%%%%%%%%%%%%%%%%%%%%%%%%%%%%%%%%%%%%%          THREE CHANNELS       %%%%%%%%%%%%%%%%%%%%%%%%%%%%%%%%%%%%%%%%%%%%%%%%%%%%%%%%%%%%%%%%%%%%%%%%%%%%%%%%%%%%%%%%%%%%%%%%%%%%%%%%%%%%%
%\begin{quote}

\smallskip
%\noindent
\textbf{Group 1 : (0,0),(0,1),(0,2)}
\smallskip

\noindent
These channels contain only two types of interactions $(b_{0,\gamma},f_{0,\gamma})$ because of the restriction in momentum conservation.
 Backscattering $b_{0,\gamma}$ exchanges the Fermi points of the incident electrons and forward scattering $f_{0,\gamma}$ does not in contrast.

\smallskip
%\noindent
\textbf{Group 2 : (1,1),(2,2)}
\smallskip

\noindent
These channels contain backward processes $(g_{1}^{1},g_{2}^{1})$ in the circumferential direction, which exchange angular momentum between electrons.
 We will consider six interactions $(g_1^1,g_1^2,g_1^4,g_2^1,g_2^2,g_2^4)$ here. 

\smallskip
\textbf{Group 3 : (1,2)}
\smallskip

\noindent
If $|n_1|+|n_2|=5$, these channels contain Umklapp processes $(g_{1}^{3},g_{2}^{3})$, which conserve the total momentum along the axial direction but does not conserve the total angular momentum along the circumferential direction.
 We have to allow the angular momentum change by a reciprocal lattice vector. 
We will consider six interactions represented by $(g_1^1,g_1^3,g_1^4,g_2^2,g_2^3,g_2^4)$. 
If $|n_1|+|n_2|\neq5$, then we shall put $g_1^3=g_2^3=0$.
%%%%%%%%%%%%%%%%%%%%%%%%%%%%%%%%%%%%%%%%%%%%%%%%%%%%%%%%%%%%%%%%%%%%%%%%%%%%%%%%%%%%%               SCALING EQUATIONS                   %%%%%%%%%%%%%%%%%%%%%%%%%%%%%%%%%%%%%%%%%%%%%%%%%%%%%%%%%%%%%%%%%%%%%%%%%%%%%%%%%%%%%%%%%%%%%%%%%%%%%%%%%%%%%%%%%%%%%%%%%%%%%%%%%%%%%%%%%%%%%%%%%%%%%%%%%%%%%%%%%%%%%
\section{Scaling equations}
Our RG calculations are straightforward generalization of the method of S\'olyom~\cite{rf:15} and performed by summing up all one-loop diagrams i.e. Cooper-channels and Peierls-channels. 
\subsection{Group 1}
These channels have two types of scattering processes ($b$; backward) and ($f$; forward). 
RG analysis of these channels is the same as pure 1D system. 
RG equations for coupling constants are 

\begin{eqnarray}
(\tilde{b}_{0,\gamma})^{\prime} &=& 2\tilde{b}^2_{0,\gamma}, \nonumber \\
(\tilde{f}_{0,\gamma})^{\prime} &=& \tilde{b}^2_{0,\gamma}.
\end{eqnarray}
Here $\tilde{b}_{0,\gamma}=b_{0,\gamma}/2 \pi v_{0,\gamma}$ and $\tilde{f}_{0,\gamma}=f_{0,\gamma}/2 \pi v_{0,\gamma}$ are normalized dimensionless coupling constants where $v_{0,\gamma} \equiv (v_{F,0}+v_{F,\gamma})/2$. 
The prime denotes differentiation with respect to $x$, where $x \equiv \ln  \frac{T}{E_{c}}$  with $E_{c}$ being an energy cutoff.
In these channels four types of TPC are produced.
 They are
\begin{eqnarray}
%%%%%%%%%%%%%%%%%%%%%%%%%%%%%%%%%%%%%%%%%%%%%%%%%%%%%%%%%%%%%%%%%%%%%%%%
\hat{O}_{0,\gamma,l}^{{\rm cdw}}(q_{1}^{+}) &=& \frac{1}{\sqrt{L}}
\sum_{k,\sigma=\pm}
{\psi^{\dagger}_{\gamma,-,l,\sigma}(k)} 
{\psi_{0,+,\sigma}(k+q_{1}^{+})}, \nonumber \\
%%%%%%%%%%%%%%%%%%%%%%%%%%%%%%%%%%%%%%%%%%%%%%%%%%%%%%%%%%%%%%%%%%%%%%%%
\hat{O}_{0,\gamma,l}^{{\rm sdw}}(q_{1}^{+}) &=& \frac{1}{\sqrt{L}}
\sum_{k,\sigma=\pm}
{\sigma \psi^{\dagger}_{\gamma,-,l,\sigma}(k)} 
{\psi_{0,+,\sigma}(k+q_{1}^{+})}, \nonumber  \\
%%%%%%%%%%%%%%%%%%%%%%%%%%%%%%%%%%%%%%%%%%%%%%%%%%%%%%%%%%%%%%%%%%%%%%%%
\hat{O}_{0,\gamma,l}^{{\rm ssc}}(q_{1}^{-}) &=& \frac{1}{\sqrt{L}}
\sum_{k,\sigma=\pm}
{\sigma \psi_{\gamma,-,l,\sigma}(k)} 
{\psi_{0,+,\sigma}(q_{1}^{-}-k)}, \nonumber \\
%%%%%%%%%%%%%%%%%%%%%%%%%%%%%%%%%%%%%%%%%%%%%%%%%%%%%%%%%%%%%%%%%%%%%%%%
\hat{O}_{0,\gamma,l}^{{\rm tsc}}(q_{1}^{-}) &=& \frac{1}{\sqrt{L}}
\sum_{k,\sigma=\pm}
{\psi_{\gamma,-,l,\sigma}(k)} 
{\psi_{0,+,\sigma}(q_{1}^{-}-k)},
%%%%%%%%%%%%%%%%%%%%%%%%%%%%%%%%%%%%%%%%%%%%%%%%%%%%%%%%%%%%%%%%%%%%%%%%
\end{eqnarray}
%%%%%%%%%%%%%%%%%%%%%%%%%%%%%%%%%%%%%%%%%%%%%%%%%%%%%%%%%%%%%%%%%%%%%%%%
where $q_{1}^{\pm} \equiv k_{F,0} \pm k_{F,\gamma}$ represents the total momentum of the TPC.
 $\rm{cdw,sdw,ssc}$ and $\rm{tsc}$ stands for charge-density wave, spin-density wave, singlet supercontuctor and triplet superconductor, respectively.
 We can also derive scaling equations for these two-particle correlation functions $\tilde{\chi}_{m}$ where $\chi _{m} \equiv \langle \hat{O}_{m}^{\dagger} \hat{O}_{m} \rangle$ and $\tilde{\chi}_{m} \equiv \pi v_{m} \frac{d}{dx} \chi_{m}$ and $m$ stands for a kind of TPC.
%%%%%%%%%%%%%%%%%%%%%%%%%%%%%%%%%%%%%%%%%%%%%%%%%%%%%%%%%%%%%%%%%%%%%%%%%
%%%%    Scaling eq for correlation functions   %%%%%%%%%%%%%%%%%%%%%%%%%%
%%%%%%%%%%%%%%%%%%%%%%%%%%%%%%%%%%%%%%%%%%%%%%%%%%%%%%%%%%%%%%%%%%%%%%%%%
 \begin{eqnarray}
 \frac{d}{dx}\log \tilde{\chi}_{m}&=&  K_{m}, \nonumber \\
 K^{0,\gamma}_{\rm{cdw}}&=&4\tilde{b}_{0,\gamma} -2\tilde{f}_{0,\gamma}, \nonumber \\
 K^{0,\gamma}_{{\rm sdw}}&=&-2\tilde{f}_{0,\gamma}, \nonumber \\
 K^{0,\gamma}_{{\rm ssc}}&=&2\tilde{b}_{0,\gamma} +2\tilde{f}_{0,\gamma}, \nonumber \\
 K^{0,\gamma}_{{\rm tsc}}&=&-2\tilde{b}_{0,\gamma} +2\tilde{f}_{0,\gamma}.
\end{eqnarray} 
 We determine the most divergent TPC by solving these scaling equations. Phase diagram for group 1 is the same as that of a pure 1D system with backward and forward scatterings as shown in Fig.2.
 %%%%%%%%%%%%%%%%%%%%%%%%%%%%%%%%%%%%%%%%%%%%%%%%%%%%%%%%%%%%%%%%%%%%%%%%%%%%%%%%%%%%%%%%%%%%%%%%%%  group1 (phase diagram)   %%%%%%%%%%%%%%%%%%%%%%%%%%%%%%%%%%%%%%%%%%%%%%%%%%%%%%%%%%%%%%%%%%%%%%%%%%%%%%%%%%%%%%%%%%%%%%%%%%%%%%%%%%%%%%
\begin{figure}[htbp]
\includegraphics[height=3.5cm,width=3.5cm]{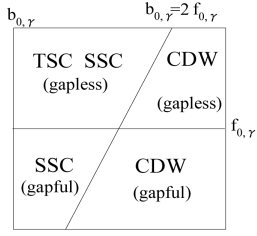}
\caption{\label{fig:epsart}
Phase diagram for group 1 in ($f_{0,\gamma},b_{0,\gamma}$) plane. 
Phase boundary is given by two lines $b_{0,\gamma}=2f_{0,\gamma}$ 
and $b_{0,\gamma}=0$.}
\end{figure}
%%%%%%%%%%%%%%%%%%%%%%%%%%%%%%%%%%%%%%%%%%%%%%%%%%%%%%%%%%%%%%%%%%%%%%%%%%%%%
 When we have $b_{0,\gamma}<0$ initially, it is scaled towards negative large quantity and diverges at finite temperature.
   We call this region gapful because gap opens in the excitation spectrum at this temperature.
%%%%%%%%%%%%%%%%%%%%%%%%%%%%%%%%%%%%%%%%%%%%%%%%%%%%%%%%%%%%%%%%%%%%%%%%%%%%%%%%%%%%%%%%%%%%%%%%%%%%%%%%%%%%%%%%%%%%%%%%%%%%%%%%%%%%%%%%%%%%%%%%%%%%%%%%%%%%%%%
\subsection{Group 2}
These channels have six scattering processes shown in Fig.3.
 Scaling equations for these six couplings are
 
%%%%%%%%%%%%%%%%%%%%%%%%%%%%%%%%%%%%%%%%%%%%%%%%%%%%%%%%%%%%%%%%%%%%%%%%%%%%%%%%%%%%%%%%%%%%%%%%%%%%scaling equations for 2ch couplings   %%%%%%%%%%%%%%%%%%%%%%%%%%%%%%%%%%%%%%%%%%%%%%%%%%%%%%%%%%%%%%%%%%%%%%%%%%%%%%%%%%%%%%%%%%%%%%%%%%%%
\begin{eqnarray}
(\tilde{g}_{1}^{1})^{\prime}&=&2(\tilde{g}_{1}^{1})^{2}+2\tilde{g}_{1}^{2}\tilde{g}_{2}^{1}, \nonumber \\
(\tilde{g}_{1}^{2})^{\prime}&=&(4\tilde{g}_{1}^{4}+2\tilde{g}_{2}^{2}-2\tilde{g}_{2}^{4})\tilde{g}_{1}^{2}+(2\tilde{g}_{1}^{1}-2\tilde{g}_{1}^{4})\tilde{g}_{2}^{1}, \nonumber \\
(\tilde{g}_{1}^{4})^{\prime}&=&2(\tilde{g}_{1}^{2})^{2}+2(\tilde{g}_{1}^{4})^{2}-2\tilde{g}_{1}^{2}\tilde{g}_{2}^{1}, \nonumber \\
(\tilde{g}_{2}^{1})^{\prime}&=&2\tilde{g}_{1}^{1}\tilde{g}_{1}^{2}+2(\tilde{g}_{2}^{2}-\tilde{g}_{2}^{4})\tilde{g}_{2}^{1}, \nonumber \\
(\tilde{g}_{2}^{2})^{\prime}&=&(\tilde{g}_{1}^{1})^{2}+(\tilde{g}_{1}^{2})^{2}+\tilde{g}_{1}^{2}\tilde{g}_{2}^{1}, \nonumber \\
(\tilde{g}_{2}^{4})^{\prime}&=&(\tilde{g}_{1}^{4})^{2}-(\tilde{g}_{2}^{1})^{2},
\end{eqnarray}
%%%%%%%%%%%%%%%%%%%%%%%%%%%%%%%%%%%%%%%%%%%%%%%%%%%%%%%%%%%%%%%%%%%%%%%%%%%%%%%%%%%%%%%%%%%%%%%%%%%%%%%%%%%%%%%%%%%%%%%%%%%%%%%%%%%%%%%%%%%%%%%%%%%%%%%%%%%%%%%
where $\tilde{g}_{i}^{j}=g_{i}^{j}/{2 \pi v_{F,\gamma}}$ are normalized coupling constants.
 Couplings are scaled towards large values and diverge at some finite temperature except for the case of $\tilde{g}_{1}^{1}=\tilde{g}_{2}^{1}=0$. 
 This suggests that the channel from group 2 tends to produce gapful phase when it contains interactions with angular momentum transfer.
 
 There are several types of TPC in these channels and  these orders have either zero or $\pm 2n_{\gamma}$ total angular momentum. Symmetry in spin space and in circumferential direction  characterize each TPC;
  either of symmetric or antisymmetric order with respect to inversion will appear both in the spin  and the circumferential direction.
 We denote all the twelve types of TPC as;
%%%%%%%%%%%%%%%%%%%%%%%%%%%%%%%%%%%%%%%%%%%%%%%%%%%%%%%%%%%%%%%%%%%%%%%%%%%%%%%%%%%%%%%%% TPC group2  %%%%%%%%%%%%%%%%%%%%%%%%%%%%%%%%%%%%%%%%%%%%%%%%%%%%%%%%%%%%%%%%%%%%%%%%%%%%%%%%%%%%%%%%%%%%%%%%%%%%%%%%%%%%%%%%%%%%%%%%%%%%%%%%%%%%%%%
\begin{widetext}
\begin{eqnarray}
%%%%%%%%%%%%%%%%%%%%%%%%%%%%%%%%%%%%%%%%%%%%%%%%%%%%%%%%%%%%%%%%%%%%%%%%
&&\hat{O}_{\gamma,\mu_{1},\mu_{2}}^{dw}(2k_{F,\gamma},0) = \frac{1}{\sqrt{2L}}
\sum_{k,\sigma,\sigma',l,l'}
{\psi^{\dagger}_{\gamma,-,l,\sigma}(k)} \ (\sigma_{\mu_{1}}^{\sigma,\sigma'}
\otimes M_{\mu_{2}}^{l,l'}) \ 
{\psi_{\gamma,+,l',\sigma'}(k+2k_{F,\gamma})}, \nonumber \\
%%%%%%%%%%%%%%%%%%%%%%%%%%%%%%%%%%%%%%%%%%%%%%%%%%%%%%%%%%%%%%%%%%%%%%%%
&&\hat{O}_{\gamma,\mu_{1}}^{dw}(2k_{F,\gamma},2l\text{L}_{\gamma}) = \frac{1}{\sqrt{L}}
\sum_{k,\sigma,\sigma'}
{\psi^{\dagger}_{\gamma,-,l,\sigma}(k)} \ \sigma_{\mu_{1}}^{\sigma,\sigma'} \ 
{\psi_{\gamma,+,-l,\sigma'}(k+2k_{F,\gamma})}, \nonumber \\
%%%%%%%%%%%%%%%%%%%%%%%%%%%%%%%%%%%%%%%%%%%%%%%%%%%%%%%%%%%%%%%%%%%%%%%%
&&\hat{O}_{\gamma,\mu_{1}}^{sc}(0,2l\text{L}_{\gamma}) = \frac{1}{\sqrt{L}}
\sum_{k,\sigma,\sigma'}
{\psi_{\gamma,-,l,\sigma}(-k)} \ \tilde{\sigma}_{\mu_{1}}^{\sigma,\sigma'} \ 
{\psi_{\gamma,+,l,\sigma'}(k)}, \nonumber \\
%%%%%%%%%%%%%%%%%%%%%%%%%%%%%%%%%%%%%%%%%%%%%%%%%%%%%%%%%%%%%%%%%%%%%%%%
&&\hat{O}_{\gamma,\mu_{1},\mu_{2}}^{sc}(0,0) = \frac{1}{\sqrt{2L}}
\sum_{k,\sigma,\sigma',l,l'}
{\psi_{\gamma,-,l,\sigma}(-k)} \ (\tilde{\sigma}_{\mu_{1}}^{\sigma,\sigma'} 
\otimes \tilde{M}_{\mu_{2}}^{l,l'}) \  
{\psi_{\gamma,+,l',\sigma'}(k)},  \\
%%%%%%%%%%%%%%%%%%%%%%%%%%%%%%%%%%%%%%%%%%%%%%%%%%%%%%%%%%%%%%%%%%%%%%%%
&&
\hspace{10mm}
 \mu_{1},\mu_{2}=\pm, \ \  \text{L}_{\gamma}=\hbar n_{\gamma}, \ \ \sigma_{\pm},M_{\pm}=
\left(
\begin{array}{cc}
1 & 0 \\
0 & \pm1
\end{array}
\right),
 \ \ \ 
 \tilde{\sigma}_{\pm},\tilde{M}_{\pm}=
\left(
\begin{array}{cc}
0 & 1 \\
\pm1 & 0
\end{array}
\right).
\nonumber
%%%%%%%%%%%%%%%%%%%%%%%%%%%%%%%%%%%%%%%%%%%%%%%%%%%%%%%%%%%%%%%%%%%%%%%%
\end{eqnarray}
\end{widetext}
%%%%%%%%%%%%%%%%%%%%%%%%%%%%%%%%%%%%%%%%%%%%%%%%%%%%%%%%%%%%%%%%%%%%%%%%
 We have abbreviated density wave and superconductor as $dw$ and $sc$ in the superscript.
 $\mu_{1}$ represents symmetry in spin space and $\mu_{2}$ represents symmetry with respect to circumferential direction.
 We use $+$ for symmetric orders and $-$ for antisymmetric orders.
 Therefore $\mu_{1}=+$ means CDW and TSC and $\mu_{1}=-$ represents SDW and SSC.
Scaling equations for these two-particle correlation functions have the same form as the first line in equations (5).
 $K_{i}$ for each orders can be calculated as
\begin{eqnarray}
%%%%%%%%%%%%%%%%%%%%%%%%%%%%%%%%%%%%%%%%%%%%%%%%%%%%%%%%%%%%%%%%%%%%%%%%%%%%%%%
&&K^{{\rm cdw}}_{\mu_{2}}(0)=(4\tilde{g}_{1}^{4}-2\tilde{g}_{2}^{4})+\mu_{2}(4\tilde{g}_{1}^{2}-2\tilde{g}_{2}^{1}), \nonumber \\
%%%%%%%%%%%%%%%%%%%%%%%%%%%%%%%%%%%%%%%%%%%%%%%%%%%%%%%%%%%%%%%%%%%%%%%%%%%%%%%
&&K^{{\rm sdw}}_{\mu_{2}}(0)=-2(\tilde{g}_{2}^{4}+\mu_{2}\tilde{g}_{2}^{1}), \nonumber \\
%%%%%%%%%%%%%%%%%%%%%%%%%%%%%%%%%%%%%%%%%%%%%%%%%%%%%%%%%%%%%%%%%%%%%%%%%%%%%%%
&&K^{{\rm cdw}} (2ln_{\gamma}) = 4 \tilde{g}_{1}^{1}-2 \tilde{g}_{2}^{2}, \nonumber \\
%%%%%%%%%%%%%%%%%%%%%%%%%%%%%%%%%%%%%%%%%%%%%%%%%%%%%%%%%%%%%%%%%%%%%%%%%%%%%%%
&&K^{{\rm sdw}}(2ln_{\gamma})=-2\tilde{g}_{2}^{2}, \nonumber 
\end{eqnarray}
\begin{eqnarray}
%%%%%%%%%%%%%%%%%%%%%%%%%%%%%%%%%%%%%%%%%%%%%%%%%%%%%%%%%%%%%%%%%%%%%%%%%%%%%%%
&&K^{{\rm ssc}}(2ln_{\gamma})=2\tilde{g}_{1}^{4}+2\tilde{g}_{2}^{4}, \nonumber \\ 
%%%%%%%%%%%%%%%%%%%%%%%%%%%%%%%%%%%%%%%%%%%%%%%%%%%%%%%%%%%%%%%%%%%%%%%%%%%%%%%
&&K^{{\rm tsc}}(2ln_{\gamma})=-2\tilde{g}_{1}^{4}+2\tilde{g}_{2}^{4}, \nonumber \\
%%%%%%%%%%%%%%%%%%%%%%%%%%%%%%%%%%%%%%%%%%%%%%%%%%%%%%%%%%%%%%%%%%%%%%%%%%%%%%%
&&K^{{\rm ssc}}_{\mu_{2}}(0)=(2\tilde{g}_{1}^{1}+2\tilde{g}_{2}^{2})+\mu_{2}(2\tilde{g}_{1}^{2}+2\tilde{g}_{2}^{1}), \nonumber \\
%%%%%%%%%%%%%%%%%%%%%%%%%%%%%%%%%%%%%%%%%%%%%%%%%%%%%%%%%%%%%%%%%%%%%%%%%%%%%%%
&&K^{{\rm tsc}}_{\mu_{2}}(0)=(-2\tilde{g}_{1}^{1}+2\tilde{g}_{2}^{2})+\mu_{2}(-2\tilde{g}_{1}^{2}+2\tilde{g}_{2}^{1}). 
%%%%%%%%%%%%%%%%%%%%%%%%%%%%%%%%%%%%%%%%%%%%%%%%%%%%%%%%%%%%%%%%%%%%%%%%%%%%%%%
\end{eqnarray}

By substituting the solution for equations (6) into (8), we obtain the temperature dependences of two-particle correlation functions. We then obtain the phase diagram in coupling space by specifying the most divergent correlation function.
%%%%%%%%%%%%%%%%%%%%%%%%%%%%%%%%%%%%%%%%%%%%%%%%%%%%%%%%%%%%%%%%%%%%%%%%%%%%%%%%%%%%%%%%%%%%%%%%%%%   About the solution for RG eq (2ch)%%%%%%%%%%%%%%%%%%%%%%%%%%%%%%%%%%%%%%%%%%%%%%%%%%%%%%%%%%%%%%%%%%%%%%%%%%%%%%%%%%%%%%%%%%%%%%%%%%%%%%%
 The phase diagrams are shown in Fig.4. 
If we assume bare value for couplings which depends only on the momentum transfer, we can put ``$\tilde{g}_{1}^{2}=\tilde{g}_{1}^{4} \equiv b$, $\tilde{g}_{2}^{2}=\tilde{g}_{2}^{4} \equiv f$'' and get four independent parameters ``$b,f,\tilde{g}_{1}^{1},\tilde{g}_{2}^{1}$''.

For the special case ``$\tilde{g}_{1}^{1}=\tilde{g}_{2}^{1}=0$'', we can solve the equation (6) exactly, 
and the solution is
\begin{eqnarray}
&&\tilde{b}(x)=\frac{\tilde{b}(0)}{1-4\tilde{b}(0)x}, \nonumber \\
&&\tilde{b}(x)-4\tilde{f}(x)=\tilde{b}(0)-4\tilde{f}(0).
\end{eqnarray}
When $\tilde{b}(0)<0$, $\tilde{b}$ diverges at finite temperature $T_{c}=E_{c} \exp(1/4\tilde{b}(0))$.
In one dimension, such an anomaly at finite temperature is an artifact of one-loop calculation, and higher-order terms will shift it to $T=0$.
 Nevertheless, this $T_{c}$ suggests us the temperature where the system crosses over from weak coupling to strong coupling.
 
 The phase diagram for this special case is shown in Fig.4(a).
 In gapless phase, all the DWs and SCs are degenerate and have the same temperature dependence of the correlation function. 
 The DW phase and the SC phase are separated by a phase boundary $b-4f=0$.
 In the gapful phase, only $\text{CDW}_{+}(0)$ phase appears, where $+$ and $(0)$ represent the symmetry $\mu_{2}=+$ and angular momentum $\text{L}=0$ respectively. 

In general, $\tilde{g}_{1}^{1}$ and $\tilde{g}_{2}^{1}$ take finite values, and other gapful phases enter into the phase diagram as shown in Fig.4(b).
 The effect of $\tilde{f}$ on the phase diagram is negligible.
Independent of the values of $\tilde{b}$ and $\tilde{f}$, there exist four gapful phases in which the most divergent order of each is $\text{SSC}_{+}(0)$, $\text{SSC}_{-}(0)$, $\text{CDW}_{+}(0)$ and $\text{CDW}_{-}(0)$ respectively. 
There is no gapless phase except for the original point $(\tilde{g}_{1}^{1},\tilde{g}_{2}^{1})=(0,0)$ even when $\tilde{b}>0$.
 Four gapful phases are characterized by asymptotic behavior of the couplings near the critical temperature $T_{c}$ ~\cite{rf:16}, which are given by
 %%%%%%%%%%%%%%%%%%%%%%%%%%%%%%%%%%%%%%%%%%%%%%%%%%%%%%%%%%%%%%%%%%%%%%%%%%%%%%%%%%%%%%%%%%%%%%%%%%% figure sanran 2ch %%%%%%%%%%%%%%%%%%%%%%%%%%%%%%%%%%%%%%%%%%%%%%%%%%%%%%%%%%%%%%%%%%%%%%%%%%%%%%%%%%%%%%%%%%%%%%%%%%%%%%%%%%%%%%%%%%%%%%
\begin{figure}[h]
\includegraphics[height=7.8cm,width=8cm]{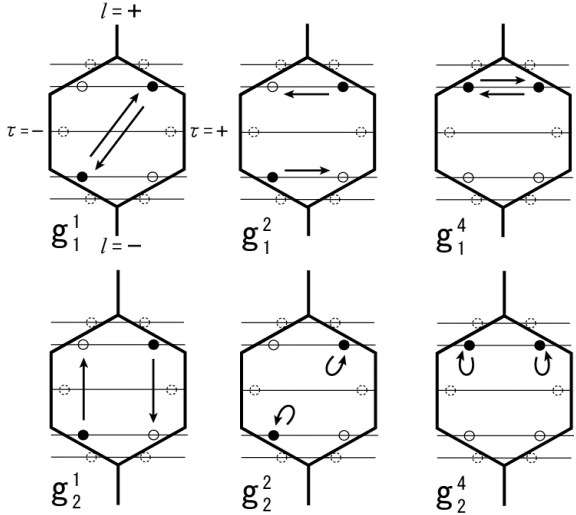}
\caption{\label{fig:epsart}
Six scattering processes in channels of group 2 are shown in momentum space when $(n_{0},n_{1},n_{2})=(0,3,2)$. 
 Hexagon represents the Brillouin zone. }
\end{figure}
\begin{eqnarray}
\tilde{g}_{i}^{j}&=&g_{i}^{j \ast}\Lambda/[1+\Lambda x],  \nonumber \\
T_{c}&=&E_{c}\exp(-\Lambda ^{-1}).
\end{eqnarray}
All the couplings are proportional to $\Lambda/[1+\Lambda x]$, which diverges at $T_{c}$ given above.
 Sets of $g_{i}^{j \ast}$ are universal for each phase as in~\cite{rf:16}, and listed in Table \ref{tab:table1}.
 Equations (6) are invariant to transformation $(\tilde{g}_{1}^{2},\tilde{g}_{2}^{1})$$\to$$(-\tilde{g}_{1}^{2},-\tilde{g}_{2}^{1})$.
 The two sets of solutions $(g_{1}^{1\ast},g_{1}^{2\ast},g_{1}^{4\ast},g_{2}^{1\ast},g_{2}^{2\ast},g_{2}^{4\ast})$ and $(g_{1}^{1\ast},-g_{1}^{2\ast},g_{1}^{4\ast},-g_{2}^{1\ast},g_{2}^{2\ast},g_{2}^{4\ast})$ correspond to two phases with different symmetry $(\mu_{2}=\pm)$ as seen in Table \ref{tab:table1}.
 The asymptotic solutions lead to behavior of TPC,
 \begin{eqnarray}
 \chi_{m} \propto |\ln \frac{T}{T_{c}}|^{\alpha_{m}+1} \propto (\frac{1}{T-T_{c}})^{-\alpha_{m}-1} \ ,
 \end{eqnarray}
where $\alpha_{m}$ determines the critical exponent for the correlation function of TPC of type $m$ and is calculated from the value of $g_{i}^{j \ast}$ using $\alpha_{m}=K_{m}|_{\tilde{g}_{i}^{j}=g_{i}^{j \ast}}$.
Correlation functions diverge at $T_{c}$ if the exponents are positive, and are suppressed if they are negative.
These exponents in each phase are listed in Table \ref{tab:table2}. \par
 It can be seen from Table \ref{tab:table1} that scattering processes with angular momentum transfer $g_{1}^{1}$ and $g_{2}^{1}$ are renormalized to 0 in the two CDW phases, on the other hand they are renormalized to large value in the two SSC phases.
  This means such scattering processes become strong attraction between electrons in a SC pair and give rise to divergence of SCC.
 We should note that all the dominant TPC in these channels carry zero angular momentum, though they have the same correlation function as TPC with angular momentum $\pm 2 \hbar n_{\gamma}$ in the non-interacting case.
 Only one type of TPC diverges and all others are suppressed in every phase.
\begin{figure}[b]
\includegraphics[height=5.1cm,width=8.3cm]{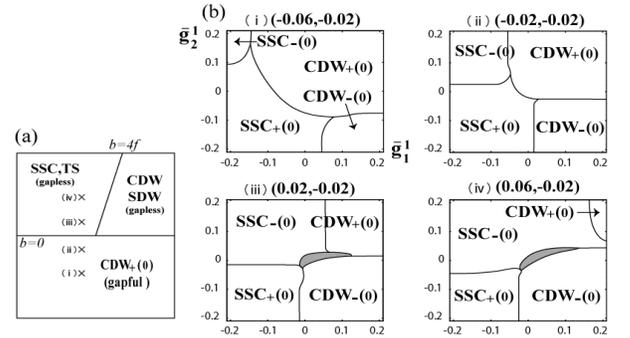}
\caption{\label{fig:epsart}
(a) Phase diagram for group 2 in ($\tilde{f},\tilde{b}$) plane is exactly obtained for the special case $\tilde{g}_{1}^{1}=\tilde{g}_{2}^{1}=0$.
(b) Phase diagram for group 2 in ($\tilde{g}_{1}^{1},\tilde{g}_{2}^{1}$) plane.
 Fixed $(\tilde{b},\tilde{f})$ values are shown above each diagram.
 In the shaded region ~\cite{rf:17}, we cannot determine which TPC is the most divergent since couplings diverge at very low temperature. }
\end{figure}
 
 We see from numerical calculations that $T_{c}$ exponentially depends on backscatterings $\tilde{b}$ and  
  $T_{c}$ becomes higher with negative large $\tilde{b}$.
 However $T_{c}$  will quickly be suppressed in case of positive $\tilde{b}$.
 Therefore phase transition can occur at 15K in these channels only when the backscatterings $\tilde{b}$ are attracting forces.
However, $T_{c}$ tends to be higher than that of group 1's for two reasons;
 (1) double degeneracy of opposite angular momenta and (2) smaller curvature of the bands $(v_{F,1} \sim v_{F,2} = 2.8 \times 10^{5} \ m/s < v_{F,0} = 6.9 \times 10^{5} \ m/s)$ will produce a larger density of states of electrons compared to that for band 1.
%%%%%%%%%%%%%%%%%%%%%%%%%%%%%%%%%%%%%%%%%%%%%%%%%%%%%%%%%%%%%%%%%%%%%%%%%%%%%%%%%%%%%%%%%%%%%%%  TABLE1 --- strong coupling parameter (group2)     %%%%%%%%%%%%%%%%%%%%%%%%%%%%%%%%%%%%%%%%%%%%%%%%%%%%%%%%%%%%%%%%%%%%%%%%%%%%%%%%%%%%%%%%%%
  \begin{table*}
\caption{\label{tab:table1}  Strong-coupling fixed-point parameters for each gapful phase in group 2}
\begin{ruledtabular}
\begin{tabular}{cccccccc}
$phase$ &$g_{1}^{1 \ast}$ & $g_{1}^{2 \ast}$ & $g_{1}^{4 \ast}$&
 $g_{2}^{1 \ast}$ & $g_{2}^{2 \ast}$ & $g_{2}^{4 \ast}$ \\
\hline
$\text{CDW}_{+}(0)$& 0 & -1/4 & -1/4 & 0
& -1/16 & -1/16 &\\
$\text{CDW}_{-}(0)$& 0 & 1/4 & -1/4 & 0
& -1/16 & -1/16 &\\
$\text{SSC}_{+}(0)$& -1/4 & -1/4 & 0 & -1/4
& -3/16 & 1/16 &\\
$\text{SSC}_{-}(0)$& -1/4 & 1/4 & 0 & 1/4
& -3/16 & 1/16 &\\
\end{tabular}
\end{ruledtabular}
\end{table*} 

%%%%%%%%%%%%%%%%%%%%%%%%%%%%%%%%%%%%%%%%%%%%%%%%%%%%%%%%%%%%%%%%%%%%%%%%%%%%%%%%%%%%%%%%%%%%%%      TABLE2 --- critical exponents (group2)           %%%%%%%%%%%%%%%%%%%%%%%%%%%%%%%%%%%%%%%%%%%%%%%%%%%%%%%%%%%%%%%%%%%%%%%%%%%%%%%%%%%%%%%%
 \begin{table*}
\caption{\label{tab:table2} Exponents $-(\alpha_{m}+1)$ in group 2 }
\begin{ruledtabular}
\begin{tabular}{cccccccc}
$phase$ &$\text{CDW}_{+}(0)$ & $\text{CDW}_{-}(0)$ & $\text{CDW}(2L_{\gamma})$&
 $\text{SDW}_{+}(0)$ & $\text{SDW}_{-}(0)$ & $\text{SDW}(2L_{\gamma})$ \\
 &$\text{SSC}_{+}(0)$ & $\text{SSC}_{-}(0)$ & $\text{SSC}(2L_{\gamma})$&
 $\text{TSC}_{+}(0)$ & $\text{TSC}_{-}(0)$ & $\text{TSC}(2L_{\gamma})$ \\
\hline
\hline
$\text{CDW}_{+}(0)$  & 7/8 & -9/8 & -9/8 & 
-9/8 & -9/8 & -9/8 &\\
                    & -3/8 & -11/8 & -3/8 &
                     -11/8 & -3/8 & -11/8 &\\
\hline
$\text{CDW}_{-}(0)$  & -9/8 & 7/8 & -9/8 & 
 -9/8 & -9/8 & -9/8 &\\
                    & -11/8 & -3/8 & -3/8 &
                     -3/8& -11/8 & -11/8 &\\
\hline
$\text{SSC}_{+}(0)$  & -3/8 & -11/8 & -3/8 &
 -11/8 & -3/8 & -11/8 &\\
                    & 7/8 & -9/8 & -9/8 &
                     -9/8 & -9/8 & -9/8 &\\
\hline
$\text{SSC}_{-}(0)$   & -11/8 & -3/8 & -3/8 &
 -3/8 & -11/8 & -11/8 &\\
                    & -9/8 & 7/8 & -9/8 &
                     -9/8 & -9/8 & -9/8 &\\
\end{tabular}
\end{ruledtabular}
\end{table*}
 
%%%%%%%%%%%%%%%%%%%%%%%%%%%%%%%%%%%%%%%%%%%%%%%%%%%%%%%%%%%%%%%%%%%%%%%%%%%%%%%%%%%%%%%%%%%%%%%%%%%%%%%   group 3   %%%%%%%%%%%%%%%%%%%%%%%%%%%%%%%%%%%%%%%%%%%%%%%%%%%%%%%
\subsection{Group 3}
This channel has six independent scattering processes shown in Fig.5.
 Scaling equations for these six couplings are 
%%%%%%%%%%%%%%%%%%%%%%%%%%%%%%%%%%%%%%%%%%%%%%%%%%%%%%%%%%%%%%%%%%%%%%%%%%%%%%%%%%%%%%%%%%%%%%%  scaling equations for group3    %%%%%%%%%%%%%%%%%%%%%%%%%%%%%%%%%%%%%%%%%%%%%%%%%%%%%%%%%%%%%%%%%%%%%%%%%%%%%%%%%%%%%%%%%%%%%%%%%%%%%%%%%%%%
 \begin{eqnarray}
(\tilde{g}_{1}^{1})^{\prime}&=&2(\tilde{g}_{1}^{1})^{2}-2\tilde{g}_{1}^{3}\tilde{g}_{2}^{3}, \nonumber \\
(\tilde{g}_{1}^{3})^{\prime}&=&(4\tilde{g}_{1}^{1}+2\tilde{g}_{2}^{4}-2\tilde{g}_{2}^{2})\tilde{g}_{1}^{3}+(2\tilde{g}_{1}^{4}-2\tilde{g}_{1}^{1})\tilde{g}_{2}^{3}, \nonumber \\
(\tilde{g}_{1}^{4})^{\prime}&=&2(\tilde{g}_{1}^{4})^{2}+2\tilde{g}_{1}^{3} \tilde{g}_{2}^{3}, \nonumber \\
(\tilde{g}_{2}^{2})^{\prime}&=&(\tilde{g}_{1}^{1})^{2}-(\tilde{g}_{2}^{3})^{2}, \nonumber \\
(\tilde{g}_{2}^{3})^{\prime}&=&2\tilde{g}_{1}^{3}\tilde{g}_{1}^{4}+(2\tilde{g}_{2}^{4}-2\tilde{g}_{2}^{2})\tilde{g}_{2}^{3}, \nonumber \\
(\tilde{g}_{2}^{4})^{\prime}&=&(\tilde{g}_{1}^{3})^{2}+(\tilde{g}_{1}^{4})^{2}+(\tilde{g}_{2}^{3})^{2}, 
\end{eqnarray}
where $\tilde{g}_{i}^{j}=g_{i}^{j}/ \pi (v_{F,1}+v_{F,2})$ in (10) are normalized scattering amplitudes. 
All couplings are scaled to large values except for the case $\tilde{g}_{1}^{3}=\tilde{g}_{2}^{3}=0$. 
 This suggests that the channel from group 3 tends to produce gapful phase when it contains Umklapp processes $\tilde{g}_{1}^{3},\tilde{g}_{2}^{3}$.
 
%%%%%%%%%%%%%%%%%%%%%%%%%%%%%%%%%%%%%%%%%%%%%%%%%%%%%%%%%%%%%%%%%%%%%%%%%%%%%%%%%%%%%%%%%%%%%%%%%%% figure sanran 3ch %%%%%%%%%%%%%%%%%%%%%%%%%%%%%%%%%%%%%%%%%%%%%%%%%%%%%%%%%%%%%%%%%%%%%%%%%%%%%%%%%%%%%%%%%%%%%%%%%%%%%%%%%%%%%%%%%%%%%%
 \begin{figure}[h]
\includegraphics[height=7.7cm,width=8cm]{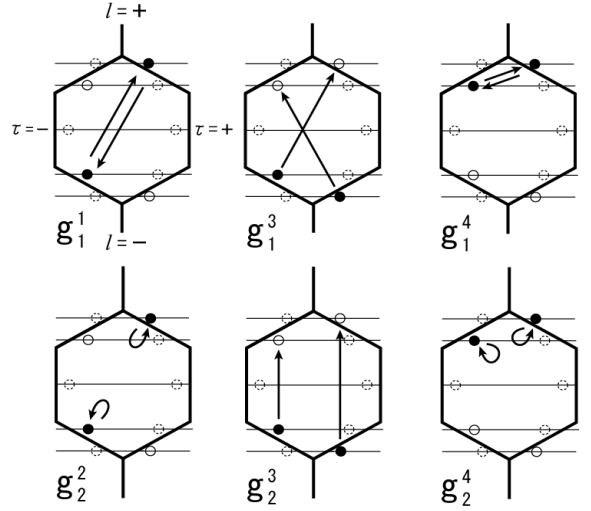}
\caption{\label{fig:epsart}
Six scattering processes in channels of group 3 are shown in momentum space when $(n_{0},n_{1},n_{2})=(0,3,2)$. 
 Hexagon represents the Brillouin zone. }
\end{figure}

There are twelve types of TPC:
%%%%%%%%%%%%%%%%%%%%%%%%%%%%%%%%%%%%%%%%%%%%%%%%%%%%%%%%%%%%%%%%%%%%%%%%%%%%%%%%%%%%%%%%% TPC group3  %%%%%%%%%%%%%%%%%%%%%%%%%%%%%%%%%%%%%%%%%%%%%%%%%%%%%%%%%%%%%%%%%%%%%%%%%%%%%%%%%%%%%%%%%%%%%%%%%%%%%%%%%%%%%%%%%%%%%%%%%%%%%%%%%%%%%%%
\begin{widetext}
\begin{eqnarray}
%%%%%%%%%%%%%%%%%%%%%%%%%%%%%%%%%%%%%%%%%%%%%%%%%%%%%%%%%%%%%%%%%%%%%%%%
&&\hat{O}_{\mu_{1}}^{dw}(q_{3}^{+},l\text{L}^{-}) = \frac{1}{\sqrt{L}}
\sum_{k,\sigma,\sigma'}
{\psi^{\dagger}_{2,-,l,\sigma}(k)} \ 
\sigma_{\mu_{1}}^{\sigma,\sigma'} \ 
{\psi_{1,+,l,\sigma'}(k+q_{3}^{+})}, \nonumber \\
%%%%%%%%%%%%%%%%%%%%%%%%%%%%%%%%%%%%%%%%%%%%%%%%%%%%%%%%%%%%%%%%%%%%%%%%
&&\hat{O}_{\mu_{1},\mu_{2}}^{dw}(q_{3}^{+},\text{L}^{+}) = \frac{1}{\sqrt{2L}}
\sum_{k,\sigma,\sigma',l,l'}
{\psi^{\dagger}_{2,-,l,\sigma}(k)} \ (\sigma_{\mu_{1}}^{\sigma,\sigma'}
\otimes \tilde{M}_{\mu_{2}}^{l,l'}) \
{\psi_{1,+,l',\sigma'}(k+q_{3}^{+})}, \nonumber \\
%%%%%%%%%%%%%%%%%%%%%%%%%%%%%%%%%%%%%%%%%%%%%%%%%%%%%%%%%%%%%%%%%%%%%%%%
&&\hat{O}_{\mu_{1}}^{sc}(q_{3}^{-},\text{L}^{+}) =  \frac{1}{\sqrt{2L}}
\sum_{k,\sigma,\sigma',l,l'}
{\psi_{2,-,l,\sigma}(q_{3}^{-}-k)} \ (\tilde{\sigma}_{\mu_{1}}^{\sigma,\sigma'} 
\otimes M_{\mu_{2}}^{l,l'}) \  
{\psi_{1,+,l',\sigma'}(k)},  \nonumber \\
%%%%%%%%%%%%%%%%%%%%%%%%%%%%%%%%%%%%%%%%%%%%%%%%%%%%%%%%%%%%%%%%%%%%%%%%
&&\hat{O}_{\mu_{1},\mu_{2}}^{sc}(q_{3}^{-},l\text{L}^{-}) =\frac{1}{\sqrt{L}}
\sum_{k,\sigma,\sigma'}
{\psi_{2,-,-l,\sigma}(q_{3}^{-}-k)} \ \tilde{\sigma}_{\mu_{1}}^{\sigma,\sigma'} \ 
{\psi_{1,+,l,\sigma'}(k)}.
% \\
%%%%%%%%%%%%%%%%%%%%%%%%%%%%%%%%%%%%%%%%%%%%%%%%%%%%%%%%%%%%%%%%%%%%%%%%
%&&
%\hspace{50mm}
%q_{3}^{\pm}=k_{F,2} \pm k_{F,3}, \ \ \text{L}^{\pm}=\hbar (n_{1} \pm n_{1}), \ %\ \ \mu_{1},\mu_{2}=\pm
%, \ \ \ \sigma_{\pm},M_{\pm}=
%\left(
%\begin{array}{cc}
%1 & 0 \\
%0 & \pm1
%\end{array}
%\right),
% \ \ \ 
% \tilde{\sigma}_{\pm},\tilde{M}_{\pm}=
%\left(
%\begin{array}{cc}
%0 & 1 \\
%\pm1 & 0
%\end{array}
%\right)
%\nonumber
%%%%%%%%%%%%%%%%%%%%%%%%%%%%%%%%%%%%%%%%%%%%%%%%%%%%%%%%%%%%%%%%%%%%%%%%
\end{eqnarray}
\end{widetext}
We use symbols $\sigma_{\mu 1},\tilde{\sigma}_{\mu 1},M_{\mu 2}$ and $\tilde{M}_{\mu 2}$ which are the same as the ones in (7).
TPC carry momentum in the axis direction $q_{3}^{\pm}=k_{F,1}\pm k_{F,2}$.
They carry angular momentum $\text{L}^{+}$ or $\text{L}^{-}$, where $\text{L}^{+}=\hbar (n_{1} + n_{2})=5\hbar$ and $\text{L}^{-}=\hbar (n_{1} - n_{2})$.
 The Umklapp processes make transition between TPC with $-\text{L}^{+}$ and TPC with $\text{L}^{+}$ in the first order correction, therefore bonding or anti-bonding states are formed.
 In other words, $\tilde{g}_{1}^{3}$ and $\tilde{g}_{2}^{3}$ produce the symmetric or anti-symmetric states with respect to the inversion of circumferential direction.

 $K_{i}$ in the scaling equations for these two-particle correlation functions are given by
\begin{eqnarray}
%%%%%%%%%%%%%%%%%%%%%%%%%%%%%%%%%%%%%%%%%%%%%%%%%%%%%%%%%%%%%%%%%%%%%%%%%%%%%%%
&&K^{{\rm cdw}} (\text{L}^{-}) = 4 \tilde{g}_{1}^{4}-2 \tilde{g}_{2}^{4}, \nonumber \\
%%%%%%%%%%%%%%%%%%%%%%%%%%%%%%%%%%%%%%%%%%%%%%%%%%%%%%%%%%%%%%%%%%%%%%%%%%%%%%%
&&K^{{\rm sdw}}(\text{L}^{-})=-2\tilde{g}_{2}^{4}, \nonumber \\
%%%%%%%%%%%%%%%%%%%%%%%%%%%%%%%%%%%%%%%%%%%%%%%%%%%%%%%%%%%%%%%%%%%%%%%%%%%%%%%
&&K^{{\rm cdw}}_{\mu_{2}}(\text{L}^{+})=(4\tilde{g}_{1}^{1}-2\tilde{g}_{2}^{2})+\mu_{2}(4\tilde{g}_{1}^{3}-2\tilde{g}_{2}^{3}), \nonumber \\
%%%%%%%%%%%%%%%%%%%%%%%%%%%%%%%%%%%%%%%%%%%%%%%%%%%%%%%%%%%%%%%%%%%%%%%%%%%%%%%
&&K^{{\rm sdw}}_{\mu_{2}}(\text{L}^{+})=-2(\tilde{g}_{2}^{2}+\mu_{2}\tilde{g}_{2}^{3}), \nonumber \\
%%%%%%%%%%%%%%%%%%%%%%%%%%%%%%%%%%%%%%%%%%%%%%%%%%%%%%%%%%%%%%%%%%%%%%%%%%%%%%%
%%%%%%%%%%%%%%%%%%%%%%%%%%%%%%%%%%%%%%%%%%%%%%%%%%%%%%%%%%%%%%%%%%%%%%%%%%%%%%%
&&K^{{\rm ssc}}_{\mu_{2}}(\text{L}^{+})=(2\tilde{g}_{1}^{4}+2\tilde{g}_{2}^{4})+\mu_{2}(2\tilde{g}_{1}^{3}+2\tilde{g}_{2}^{3}), \nonumber \\
%%%%%%%%%%%%%%%%%%%%%%%%%%%%%%%%%%%%%%%%%%%%%%%%%%%%%%%%%%%%%%%%%%%%%%%%%%%%%%%
&&K^{{\rm tsc}}_{\mu_{2}}(\text{L}^{+})=(-2\tilde{g}_{1}^{4}+2\tilde{g}_{2}^{4})+\mu_{2}(-2\tilde{g}_{1}^{3}+2\tilde{g}_{2}^{3}), \nonumber \\
%%%%%%%%%%%%%%%%%%%%%%%%%%%%%%%%%%%%%%%%%%%%%%%%%%%%%%%%%%%%%%%%%%%%%%%%%%%%%%%
&&K^{{\rm ssc}}(\text{L}^{-})=2\tilde{g}_{1}^{1}+2\tilde{g}_{2}^{2}, \nonumber \\
%%%%%%%%%%%%%%%%%%%%%%%%%%%%%%%%%%%%%%%%%%%%%%%%%%%%%%%%%%%%%%%%%%%%%%%%%%%%%%%
&&K^{{\rm tsc}}(\text{L}^{-})=-2\tilde{g}_{1}^{1}+2\tilde{g}_{2}^{2}. 
%%%%%%%%%%%%%%%%%%%%%%%%%%%%%%%%%%%%%%%%%%%%%%%%%%%%%%%%%%%%%%%%%%%%%%%%%%%
\end{eqnarray}

If we have $\tilde{g}_{1}^{3}=\tilde{g}_{2}^{3}=0$ initially, these equations can be solved and phase diagram becomes equivalent to that of group 1.
 In this case scaling equations (12) are reduced to
 \begin{eqnarray}
 (\tilde{g}_{1}^{1})'=2(\tilde{g}_{1}^{1})^{2}, \ \ \  (\tilde{g}_{2}^{2})'=(\tilde{g}_{1}^{1})^{2}, \nonumber \\
 (\tilde{g}_{1}^{4})'=2(\tilde{g}_{1}^{4})^{2}, \ \ \  (\tilde{g}_{2}^{4})'=(\tilde{g}_{1}^{4})^{2}. \nonumber 
 \end{eqnarray}
 $\tilde{g}_{1}^{1}$ and $\tilde{g}_{1}^{4}$ diverge at different temperature $T_{c}=E_{c} \exp(1/2\tilde{g}_{1}^{1})$ and $T_{c}'=E_{c} \exp (1/2\tilde{g}_{1}^{4})$ respectively.
 Gapless phase appears in the region $\tilde{g}_{1}^{1},\tilde{g}_{1}^{4}>0$, and gapful phase appears in the region $\tilde{g}_{1}^{1},\tilde{g}_{1}^{4}<0$.

By solving (12) and (14) numerically, we obtain phase diagrams (Fig.6) for the case where Umklapp processes have some finite value $(\tilde{g}_{1}^{3}, \tilde{g}_{2}^{3} \neq 0)$. 
%%%%%%%%%%%%%%%%%%%%%%%%%%%%%%%%%%%%%%%%%%%%%%%%%%%%%%%%%%%%%%%%%%%%%%%%%%%%%%%%%%%%%%%%%%%%%%%%%    phase diagram in group3      %%%%%%%%%%%%%%%%%%%%%%%%%%%%%%%%%%%%%%%%%%%%%%%%%%%%%%%%%%%%%%%%%%%%%%%%%%%%%%%%%%%%%%%%%%%%%%%%%%%%%%%%%%%
\begin{figure}[b]
\includegraphics[height=5.4cm,width=7.2cm]{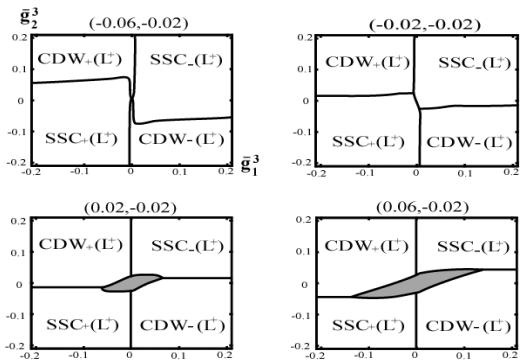}
\caption{\label{fig:epsart}
Phase diagram for group 3 in ($\tilde{g}_{1}^{3},\tilde{g}_{2}^{3}$) plane. 
Fixed $(\tilde{b},\tilde{f})$ values are shown above each diagram.
 In the shaded region, we cannot determine which TPC is the most divergent since couplings diverge only at very low temperatures, where our calculation becomes less reliable. }
\end{figure}

 We have taken $\tilde{g}_{1}^{1}=\tilde{g}_{1}^{4}=\tilde{b}$ and $\tilde{g}_{2}^{2}=\tilde{g}_{2}^{4}=\tilde{f}$ for simplicity. 
The effect of $\tilde{f}$ on the phase diagram is negligible.
 Independent of $\tilde{b},\tilde{f}$ values, phase diagram mainly consists of four large gapful region $\{\text{CDW}_{\pm}(\text{L}^{+}),\text{SSC}_{\pm}(\text{L}^{+})\}$, and there is no gapless phase when we have non-zero Umklapp processes $(\tilde{g}_{1}^{3},\tilde{g}_{2}^{3})$.
 These gapful phases are characterized by the asymptotic solutions (10) and (11) as well as in group 2.
 We show the values of $g_{i}^{j \ast}$ in Table \ref{tab:table3}, and exponents $-(\alpha_{m}+1)$ in Table \ref{tab:table4}.
  Equations (12) are invariant to transformation $(\tilde{g}_{1}^{3},\tilde{g}_{2}^{3})$$\to$$(-\tilde{g}_{1}^{3},-\tilde{g}_{2}^{3})$.
 The two solutions $(g_{1}^{1\ast},g_{1}^{3\ast},g_{1}^{4\ast},g_{2}^{2\ast},g_{2}^{3\ast},g_{2}^{4\ast})$ and $(g_{1}^{1\ast},-g_{1}^{3\ast},g_{1}^{4\ast},g_{2}^{2\ast},-g_{2}^{3\ast},g_{2}^{4\ast})$ correspond to two phases with different symmetry $(\mu_{2}=\pm)$ as seen from Table \ref{tab:table3}.

In $\text{CDW}_{\mu_{2}}(\text{L}^{+})$ phase, the correlation function for both $\text{CDW}_{\mu_{2}}(\text{L}^{+})$ and $\text{TSC}_{-\mu_{2}}(\text{L}^{+})$ diverges at $T_{c}$ and all others are suppressed.
 Correlation function for $\text{SSC}_{\mu_{2}}(\text{L}^{+})$ alone diverges in $\text{SSC}_{\mu_{2}}(\text{L}^{+})$ phase.
 Every divergent TPC carry angular momentum $\text{L}^{+}=5\hbar$.
 Umklapp processes $(\tilde{g}_{1}^{3},\tilde{g}_{2}^{3})$ are scaled towards large value in any case, which implies that they play an important role for TPC with angular momentum $\text{L}^{+}$ to develop.
 $T_{c}$ is exponentially dependent on the value of backscatterings $\tilde{b}$ as in the case of group 2.
 When $\tilde{b}>0$ $T_{c}$ is suppressed to very small energy scale.
 This seems to be a trace of gapless phase in the special case $\tilde{g}_{1}^{3}=\tilde{g}_{2}^{3}=0$.
 When $\tilde{b}<0$, however, $T_{c}$ will be higher than that of the gapful phase in group 1 for the same reason stated in group 2.
 %%%%%%%%%%%%%%%%%%%%%%%%%%%%%%%%%%%%%%%%%%%%%%%%%%%%%%%%%%%%%%%%%%%%%%%%%%%%%%%%%%%%%%%%%%%%%%%  TABLE3 --- strong coupling parameter (group3)     %%%%%%%%%%%%%%%%%%%%%%%%%%%%%%%%%%%%%%%%%%%%%%%%%%%%%%%%%%%%%%%%%%%%%%%%%%%%%%%%%%%%%%%%%%
\begin{table*}
\caption{\label{tab:table3}  Strong coupling fixed point parameters for each gapful phase in group 3}
\begin{ruledtabular}
\begin{tabular}{cccccccc}
$phase$ &$g_{1}^{1 \ast}$ & $g_{1}^{3 \ast}$ & $g_{1}^{4 \ast}$&
 $g_{2}^{2 \ast}$ & $g_{2}^{3 \ast}$ & $g_{2}^{4 \ast}$ \\
\hline
$\text{CDW}_{+}(\text{L}^{+})$& -0.1258 & -0.4219 & 0.0810 & -0.0034
& 0.1116 & -0.1970 &\\
$\text{CDW}_{-}(\text{L}^{+})$& -0.1258 & 0.4219 & -0.0810 & -0.0034
& -0.1116 & -0.1970 &\\
$\text{SSC}_{+}(\text{L}^{+})$& 0.1026 & -0.2302 & -0.2764 & 0.0616
& -0.2685 & -0.2015 &\\
$\text{SSC}_{-}(\text{L}^{+})$& 0.1026 & 0.2302 & -0.2764 & 0.0616
& 0.2685 & -0.2015 &\\
\end{tabular}
\end{ruledtabular}
\end{table*}
%%%%%%%%%%%%%%%%%%%%%%%%%%%%%%%%%%%%%%%%%%%%%%%%%%%%%%%%%%%%%%%%%%%%%%%%%%%%%%%%%%%%%%%%%%%%%%      TABLE4 --- critical exponents (group3)           %%%%%%%%%%%%%%%%%%%%%%%%%%%%%%%%%%%%%%%%%%%%%%%%%%%%%%%%%%%%%%%%%%%%%%%%%%%%%%%%%%%%%%%%
 \begin{table*}
\caption{\label{tab:table4} Exponents $-(\alpha_{m}+1)$ in group 3 }
\begin{ruledtabular}
\begin{tabular}{cccccccc}
$phase$ &$\text{CDW}_{+}(\text{L}^{+})$ & $\text{CDW}_{-}(\text{L}^{+})$ & $\text{CDW}(\text{L}^{-})$&
 $\text{SDW}_{+}(\text{L}^{+})$ & $\text{SDW}_{-}(\text{L}^{+})$ & $\text{SDW}(\text{L}^{-})$ \\
  &$\text{SSC}_{+}(\text{L}^{+})$ & $\text{SSC}_{-}(\text{L}^{+})$ & $\text{SSC}(\text{L}^{-})$&
 $\text{TSC}_{+}(\text{L}^{+})$ & $\text{TSC}_{-}(\text{L}^{+})$ & $\text{TSC}(\text{L}^{-})$ \\
\hline
\hline
$\text{CDW}_{+}(\text{L}^{+})$  & 1.4073 & -2.4143 & -1.7182 & 
-0.7836 & -1.2299 & -1.3940 &\\
                    & -0.1474 & -1.3886 & -0.7416 &
                     -1.5109 & 0.6231 & -1.2449 &\\
\hline
$\text{CDW}_{-}(\text{L}^{+})$  & -2.4143 & 1.4073 & -1.7182 & 
-1.2299 & -0.7836 & -1.3940 &\\
                    & -1.3886 & -0.1474 & -0.7416 &
                     0.6231 & -1.5109 & -1.2449 &\\
\hline
$\text{SSC}_{+}(\text{L}^{+})$  & -0.9034 & -1.6708 & -0.2973 &
 -1.4139 & -0.3398 & -1.4030 &\\
                    & 0.9531 & -1.0416 & -1.3283 &
                     -1.0732 & -1.2265 & -0.9180 &\\
\hline
$\text{SSC}_{-}(\text{L}^{+})$   & -1.6708 & -0.9034 & -0.2973 &
 -0.3398 & -1.4139 & -1.4030 &\\
                    & -1.0416 & 0.9531 & -1.3283 &
                    -1.2265 & -1.0732 & -0.9180 &\\
\end{tabular}
\end{ruledtabular}
\end{table*}

%%%%%%%%%%%%%%%%%%%%%%%%%%%%%%%%%%%%%%%%%%%%%%%%%%%%%%%%%%%%%%%%%%%%%%%%%%%%%%%%%%%%%%%%%%           Summary     %%%%%%%%%%%%%%%%%%%%%%%%%%%%%%%%%%%%%%%%%%%%%%%%%%%%%%%%%%%%%%%%%%%%%%%%%%%%%%%%%%%%%%%%%%%%%%%%%%%%%%%%%%%%%%%%%%%%%%%%%%%%%%
\section{Summary}
We have discussed possible phase transitions in (5,0) CNs, which is quasi-1D system with angular momentum, by evaluating the most divergent two-particle correlation function.

 We have found new type of superconducting phase, where a Cooper pair carries non-zero angular momentum and Umklapp processes play significant role. 
 All other dominant TPC are singlet-superconducting or charge-density wave with zero or $5\hbar$ angular momentum, which are caused by backscattering or Umklapp scattering.
 Back scattering in the circumferential direction strengthen the correlation between electrons with opposite angular momentum and cause divergent TPC with zero angular momentum.
 On the other hand Umklapp processes strengthen the correlation between electrons with angular momentum in the same direction, and cause divergent TPC with angular momentum.

  $T_{c}$ evaluated from RG equations become quite low with repulsive interactions between electrons. So we need a large reduction of Coulomb interactions or strong attractions by electron-phonon interactions for the phase transition to occur at 15K. 
 However, in group 2 and 3, $T_{c}$ will be higher than that of group 1 because they have higher density of states at Fermi level.
 Therefore it seems that group 2 and group 3 play a critical role in the superconductivity observed in (5,0) CN.
   A recent work~\cite{rf:12} has shown that acoustic phonon exchanges result in strong attractive interactions in nanotubes with smaller diameter.
 Thus, there may be some possibility of understanding the unusually high transition temperature (15K) with our result.
 
In our study we have not considered the effects of pair-tunneling processes such as $\{(\gamma,\pm\tau),(\gamma,\mp\tau)\}$ $\to$ $\{(\gamma',\pm\tau),(\gamma',\mp\tau)\}$. 
 Such processes are known to be important for the formation of SCC in $(n,n)$ CNs~\cite{rf:5,rf:7,rf:8} and in two- and three-leg ladder systems~\cite{rf:18} as well.
 Therefore it is also important to study the effects of pair-tunneling processes.
 However things will be more difficult because pair-tunneling processes will mix with the TPC in every channels and we cannot treat three groups independently. Effects of pair-tunneling processes will be discussed elsewhere.
\begin{acknowledgments}
We thank N. Hatakenaka, B.H. Valtan, S. Tuchiya and N. Yokoshi for significant comments and discussions. 
K. K is also grateful to H.M.  Yanagisawa for encouraging him a lot.
\end{acknowledgments}
%%%%%%%%%%%%%%%%%%%%%%%%%%%%%%%%%%%%%%%%%%%%%%%%%%%%%%%%%%%%%%%%%%%%%%%%%%%%%%%%%%%%%%%%%%%%%              REFERENCES                %%%%%%%%%%%%%%%%%%%%%%%%%%%%%%%%%%%%%%%%%%%%%%%%%%%%%%%%%%%%%%%%%%%%%%%%%%%%%%%%%%%%%%%%%%%%%%%%%%%%%%%%%%

\end{document}